\documentclass[12pt]{article}
\usepackage{color}

\input{tcilatex}

\begin{document}

\textbf{Non-monotonic spontaneous magnetization in a Sznajd-like Consensus
Model } \bigskip

Lorenzo Sabatelli$^{\mathrm{1}\mathrm{,} \mathrm{2}}$ and Peter Richmond$^{%
\mathrm{1}}$ \bigskip

$^{\mathrm{1}}$Department of Physics, Trinity College Dublin, Ireland

$^{\mathrm{2}}$ HIM, IFSC, Dublin, Ireland

\bigskip

E-mail: {\color{black} \underline {sabatell@tcd.ie}}

\bigskip

\noindent

\textbf{Abstract.}Ising or Potts models [1] of ferromagnetism have been
widely used to describe locally interacting social or economic
systems[2][3][4]. We consider a related model, introduced by Sznajd[5][6] to
describe the evolution of consensus in a society. In this model, the opinion
or state of any spins can only be changed through the influence of
neighbouring pairs of similarly aligned spins. Such pairs can polarize their
neighbours. We show that, assuming the global dynamics evolve in a
synchronous manner, the two-state Sznajd model exhibits a non-monotonically
decreasing overall orientation that has a maximum value when the system is
subject to a finite value of noise. Reinterpreting the model in terms of
opinions within a society we predict that consensus can be increased by the
addition of an appropriate amount of random noise. These features are
explained by the presence of islands of complete orientation that are stable
in the absence of noise but removed via the presence of added noise.

\bigskip


\bigskip

\textbf{\ Introduction }

Simple cellular automaton based models, such as the Ising or Potts models,
are useful not only to understand  physical phenomena such as ferromagnetism
[1], but also to study the effect of interactions between human beings in a
society. In this way we may cast light on the way opinions or behaviours
spread throughout society. 

We examine a cellular automaton consensus model first introduced by Sznajd.
The model is based on the idea that, within human societies, it is
generally easier to change someone's opinion by acting within from within a
group than by acting alone. To quote Abraham Lincoln: 'United we stand,
divided we fall'.

The simplest version of the Sznajd model is implemented on a two-dimension
lattice. Each site carries a spin, $S$ that may be either up or down. This
represents either a positive or negative opinions on any question. Two
neighbouring parallel spins, representing for instance two neighbouring
people sharing the same opinion, in the model, are able to convince their
neighbours of this opinion. If these two neighbours are not parallel or in
step, then they have no influence on their neighbours. Allowing such a
system to evolve from one time step to another via a random sequential
updating mechanism, always leads after a sufficiently long time to complete
orientation of spins which for the social systems is analogous to complete
consensus providing that the initial net orientation of spins is greater
than zero.

However, if the random sequential updating is replaced by a synchronous
updating mechanism the possibility of reaching total alignment is reduced
quite dramatically. Now, updating is performed by going systematically
through the lattice to find the first member of the pair, then choosing
randomly the second member of the pair within the neighbourhood of the
first. Having in this way completed the assembly of pairs, each spin then
orients itself according to its neighbours at time step $t$. Parallel pairs
of spins will induce their neighbours to turn to their same state. However a
single spin may often belong, simultaneously, to the neighbourhood of more
than one couple (of like-oriented spins). In this case, if the neighbouring
pairs have different orientations, the state of the individual spin does not
change. It is also possible to assume that each spin has a memory of its
past $T$ orientations. The introduction of such a mechanism clearly helps
overcome the frustration of individual spins [9] however we shall not
consider this aspect in this short note. 

\bigskip 

With synchronous updating only if the initial net orientation of spins is
above a critical value (that in turn depends on the lattice size $L$ [8] and
memory length $T$ [9]) does the system reach complete orientation or
complete consensus. Below the critical value, the system evolves to a
partial orientation or consensus.

This phenomenon has, thus far, been studied at zero temperature or the
absence of noise. In this note we examine the influence of noise in the two
dimensional Sznajd model where updating is synchronous.

\bigskip

\textbf{\ Synchronous updating in the Sznajd Model with added noise}

We define the Magnetization in the following way

 $M=|(N(up)-N(down))|/(N(up)+N(down))$

where $N(up)$ is the number of up spins and $N(down)$ is the number of down spins and, as we have already
remarked, neglect memory effects. During the evolution of the system from
the initial random state into a stable configuration, we allow each spin to
flip randomly with probability $q$ where $0<q<0.5$.

Figure 1 shows the variation on a log-log scale of $1-Mc$ with  lattice
size, $L$, in the absence of noise or zero temperature. When the initial net
magnetisation, $M(0)>Mc$ the system evolves to complete consensus $(M(\infty
)=1)$.  When the initial net magnetisation, $M(0)<Mc$ the system never
reaches complete consensus $(M(\infty )<1)$.

Figures 2 shows the effect of temperature or noise on the final
magnetisation, $M(\infty )$ for lattice size of 50. Figures 3 and 4 are
similar results for lattice sizes of 100 and 250 respectively.When $M(0)>Mc$
the final orientation, $M(\infty )$ decreases monotonically as $q$
increases. When $M(0)<Mc$ the final orientation $M(\infty )$ displays
non-monotonic behaviour as a function of $q$. For very small values of $q$
(typically below 0.005), $M$ increases and may reach values close (but
still below) $1$, for intermediate values of $q$ (roughly between 0.006 and
0.06) $M(\infty )$\ may decrease slowly or even display oscillations
depending on the value of the initial magnetization. For larger values of $q$%
, $M(\infty )$ quickly decays to zero. The results, captured in Figures 2, 3
and 4 for different values of lattice size $L$, have been obtained using
Monte Carlo simulations and averaging over ensembles of 500 realizations for
each set of the three parameters $\{L,M(0),q\}$.

\bigskip

\textbf{Microstructure}

To gain further insight into this non-monotonic behaviour, we began with a
randomly selected configuration of spins on the lattice, having a fraction $p
$ up and 1-$p$ down (say $p$=0.4, i.e. initial magnetization $M(0)$%
=0.2), and no added random noise ($q$=0). After only a few hundred
iterations, the system evolved into an archipelago-like structure. Spins
could be observed aligned in the manner of  islands located in a sea of
opposite orientation. For values of $p$ close to 0.5, the 'landscape' is
irregular and jagged. The geometrical structure of these small islands is
now crucial. Spins surrounded by four others of the same sign (so forming
cross-shaped islands) form a stable state since, at any time, they are
prevented, at zero temperature, from flipping by frustration. Islands made
up of one or more copies (even overlapping) of such a cross-shaped structure
are destined to last and no total orientation can be reached in the system.

However now the addition of random noise can flip a fraction $q$ of spins,
at each time step. The occasional flipping of one spin within a cross-shaped
island now removes the frustration and breaks the previous stability of the
island structure. Equally some new cross-shaped islands may also be randomly
created. $M$ thus is now dependent on the balance between creation and
destruction of cross-shaped islands. The effect is illustrated by figures 5
and 6 which represent the microstructure at two consecutive time steps. The
area located by the arrow show how an island has become destabilised by the
noise.

\bigskip

\textbf{\ Conclusions}

A non-monotonic dependence of magnetization on random noise (temperature)
arises in a square lattice model where spins interact via a synchronous
Sznajd fashion. This is due to the presence of stable 'cross-shaped' islands
of parallel spins. Random fluctuations can destabilise these islands. So,
returning to our social analogue, one might expect, if the model applies,
that consensus may actually increase when a small amount of noise is
present. These results suggest that a certain degree of individual freedom
and independence of thought may actually increase the degree of consensus
within a society.

A more detailed analysis of these effects will be subject of further study.

\bigskip

\paragraph{Acknowledgments}

\bigskip

We thank Professors Dietrich Stauffer, Charles Patterson and Mike Coey FRS
for helpful discussions. The authors also acknowledge support from the EU
via Marie Curie Industrial Fellowship MCFH-1999-000

\par Fig.1. \par The straight line (estimated slope $-0.39$) displays in log-log scale the quantity $1-Mc $ as a function of $L$, at zero temperature.
$Mc$ is the initial net magnetization at the phase transition point from the state without consensus to the state with consensus.$L$ is the lattice linear dimension. \par

\par Fig.2 \par Magnetization at equilibrium as a function of $q$, for $L$=50 and initial magnetization values 0.9,0.2,0.02 and zero. \par

\par Fig.3 \par Magnetization at equilibrium as a function of $q$, for $L$=100 and initial magnetization values 0.9,0.2,0.02 and zero. \par

\par Fig.4 \par Magnetization at equilibrium as a function of $q$, for $L$=250 and initial magnetization values 0.9,0.2,0.02 and zero. \par

\par Fig.5 \par This shows the microscopic detail of a calculation for a lattice of linear dimension $L$=50,a probability of random flipping $q$=0.001 and an initial magnetization $M(0)$=0.2.The arrow points to the 'cross-shaped' island (positive spins, black) centred in the site {33,17} and surrounded by the sea (negative spins, white) that has formed after 396 time-steps. \par

\par Fig.6 \par This shows the microscopic detail for the calculation shown in figure 5 at the subsequent time step. The random flipping of spin {33, 16} (see the arrow) has now eased the frustration, it would otherwise exhibit at zero temperature, destabilising the 'cross shaped' island. In this particular case it took just one time-step for the other four positive spins forming the island to be flipped via interaction with the surrounding negative spins. In general that may take up to tens of steps. \par


\begin{thebibliography}{9}
\bibitem{[1]}  See for example, Huang, K. Statistical Mechanics. John Wiley 

\bibitem{[2]}  See for example, Vaga, T., Profiting from Chaos. McGraw Hill
New York 1994 ISBN 0 07 066786 1

\bibitem{[3]}  Krawiecki,A.; Holyst, J.A.; Helbing D. Volatility Clustering
and Scaling for Financial Time Series due to Attractor Bubbling. Phys. rev
Lett., Vol 89, Number15 (October 2002)

\bibitem{[4]}  Holyst, J. A.;Kacperski, K.; Schweitzer, F., in Annual
Reviews of Computational Physics IX. p.275, World Scientific, Singapore 2001

\bibitem{[5]}  Sznajd-Weron, K.; Sznajd, J. Opinion Evolution in Closed
Community. Int. J. Mod. Phys. C 11 1157-1166 (2000).

\bibitem{[6]}  Stauffer, D. Monte Carlo Simulations of the Sznajd model,
Journal of Artificial Societies and Social Simulation 5, No.1 paper 4 (2002)
(jasss.soc.surrey.ac.uk).

\bibitem{[7]}  Stauffer, D.; de Oliveira, P.M.C. Persistence of opinion in
the Sznajd consensus model: computer simulation. Eur. Phys. J. B 30, 587-592
(2003).

\bibitem{[8]}  Stauffer, D. Frustration from Simultaneous Updating in Sznajd
Consensus Model. (cond-mat/0207598 Preprint for J. Math. Sociology)

\bibitem{[9]}  Sabatelli, L.; Richmond, P. Phase transitions, memory and
frustration in a Sznajd-like model with synchronous updating. Int. J. Mod.
Phys C 14 No. 9 (2003) (cond-mat/0305015).
\end{thebibliography}
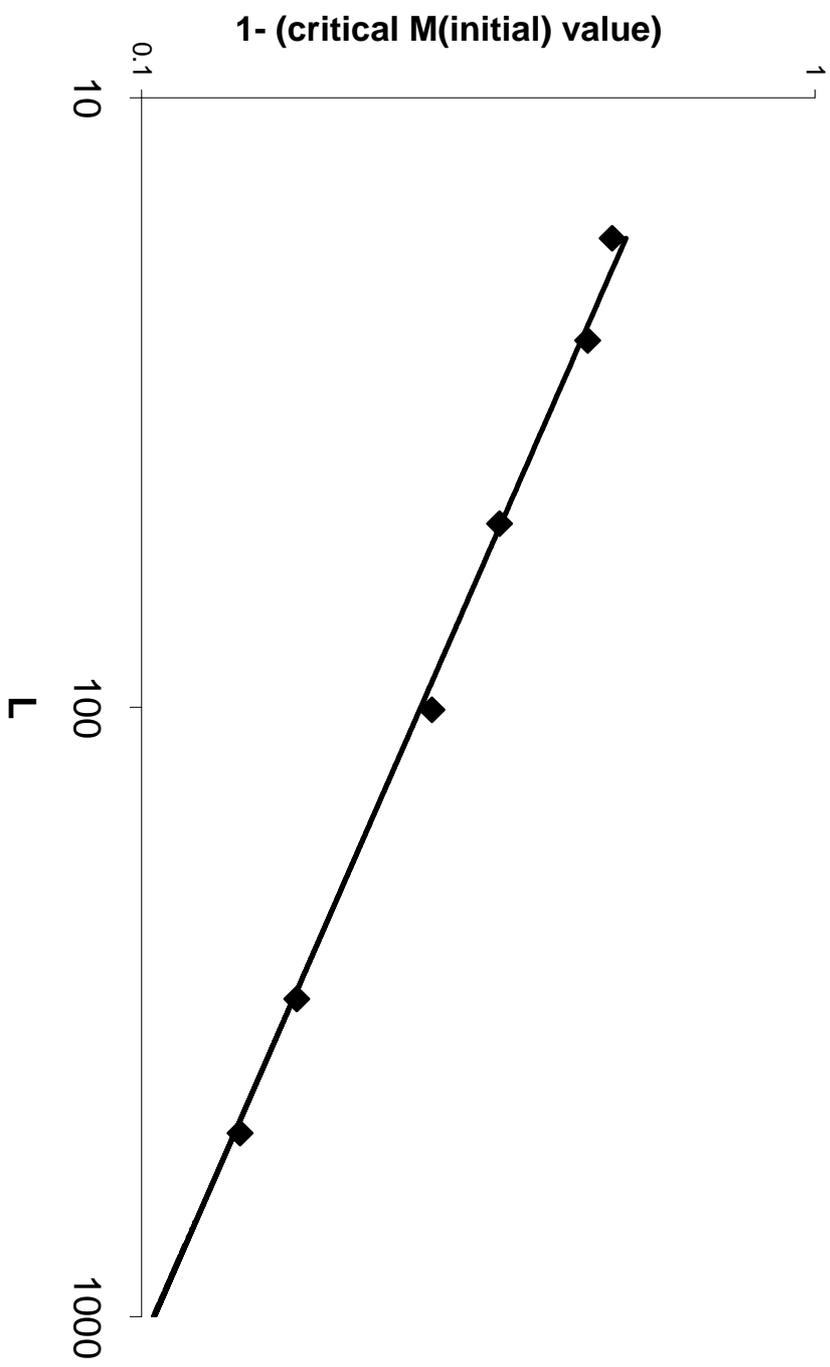
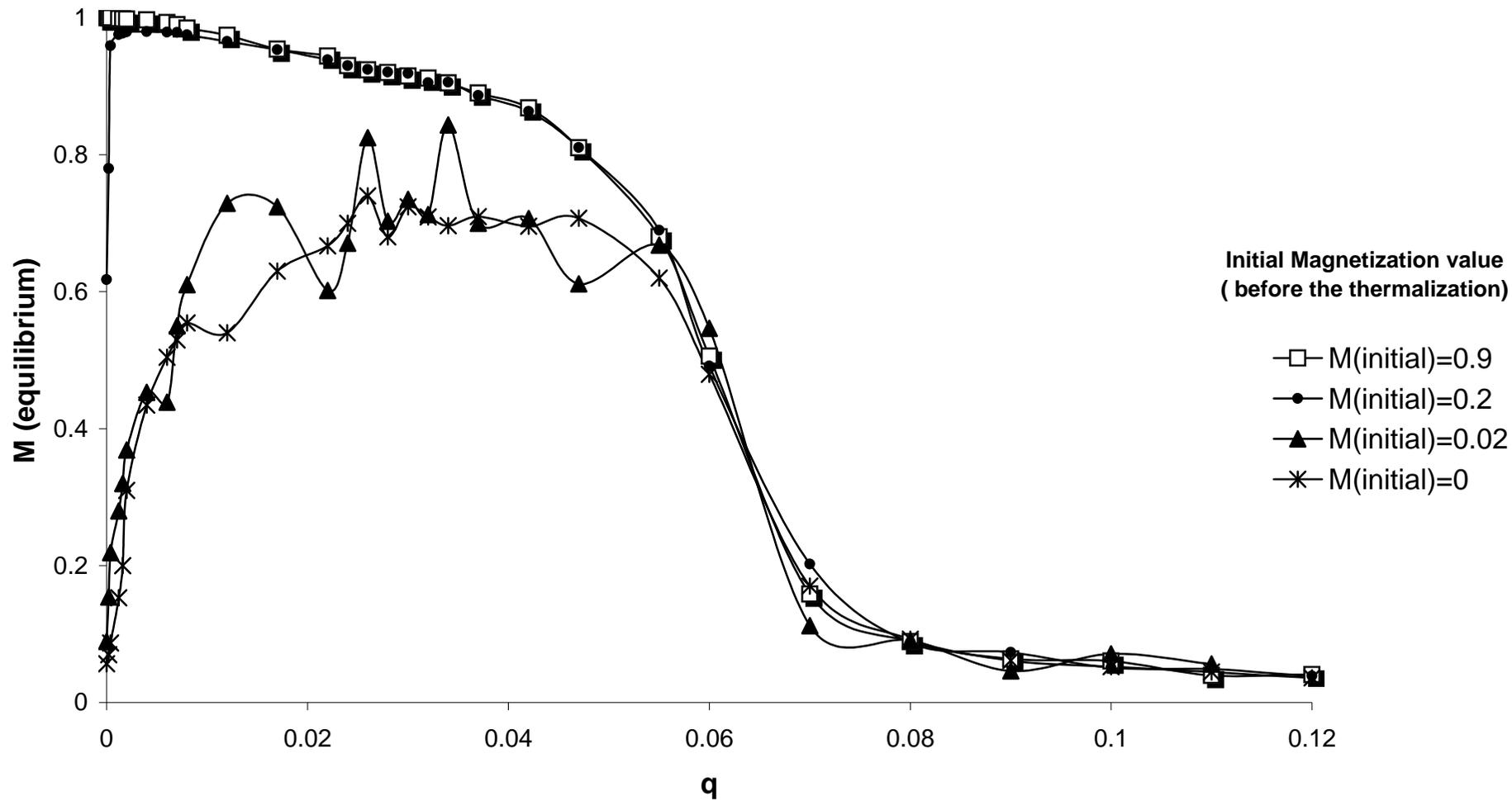
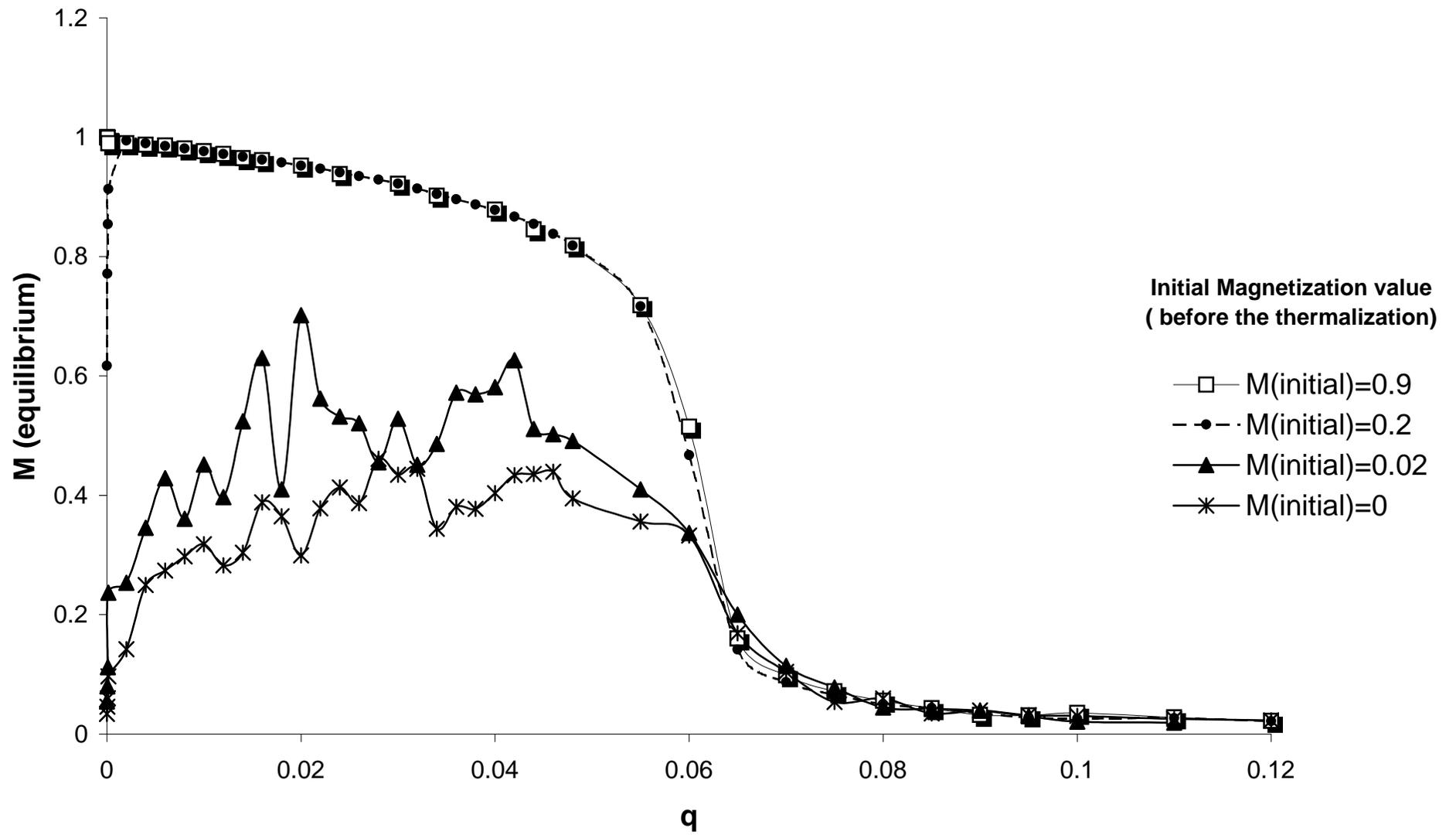
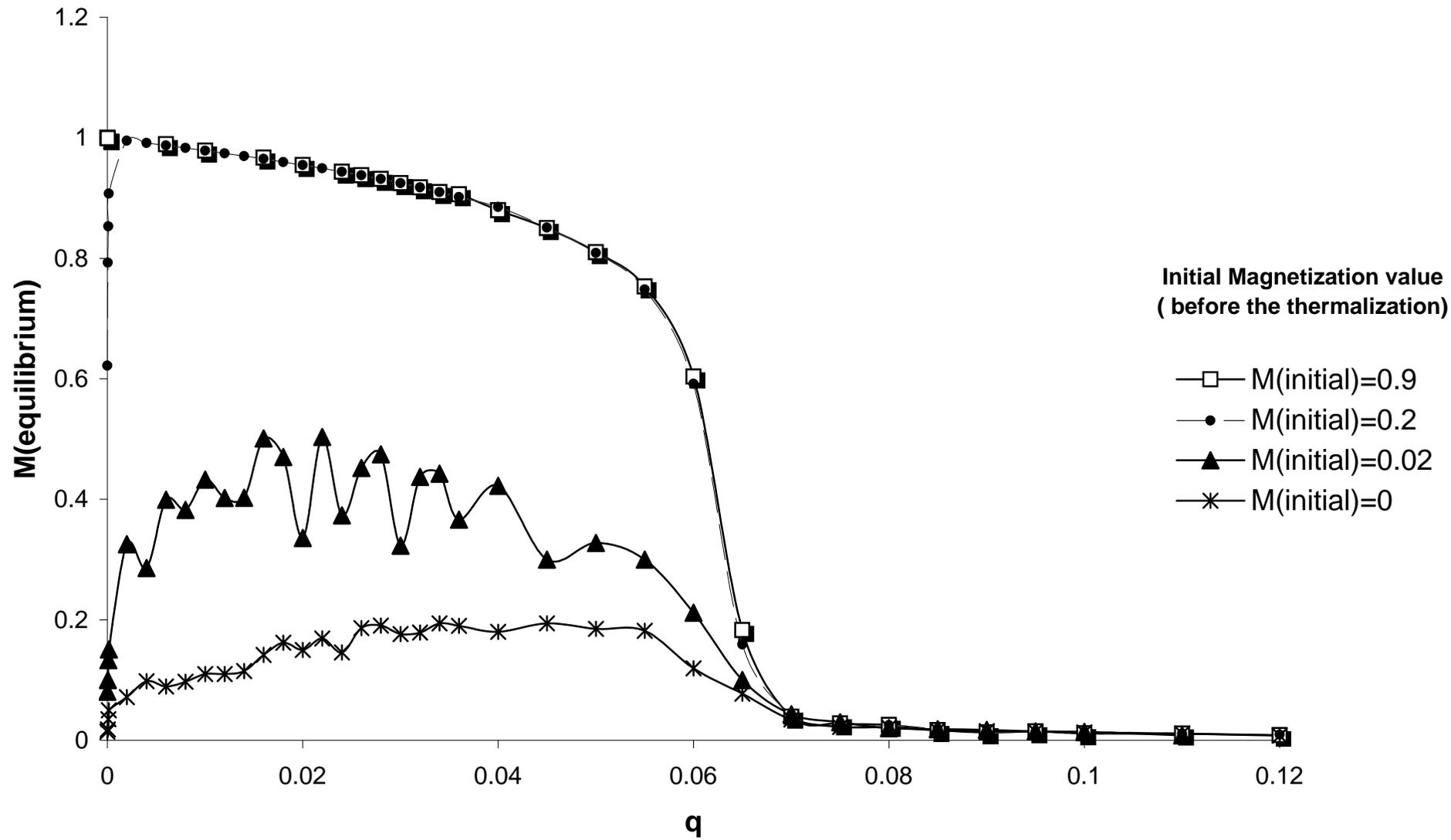
\end{document}